\documentclass[preprint,showpacs,amsmath,amssymb,amsfonts,nofootinbib]{revtex4}
\usepackage{graphicx}

\def\bef{\begin{figure}}
\def\eef{\end{figure}}

\newcommand{\bd}{\begin{displaymath}}
\newcommand{\ed}{\end{displaymath}}

\def\lsim{\raise0.3ex\hbox{$\;<$\kern-0.75em\raise-1.1ex
          e\hbox{$\sim\;$}}}
\def\gsim{\raise0.3ex\hbox{$\;>$\kern-0.75em\raise-1.1ex
\hbox{$\sim\;$}}}

\def\simlt{\mathrel{\lower2.5pt\vbox{\lineskip=0pt\baselineskip=0pt
           \hbox{$<$}\hbox{$\sim$}}}}
\def\simgt{\mathrel{\lower2.5pt\vbox{\lineskip=0pt\baselineskip=0pt
           \hbox{$>$}\hbox{$\sim$}}}}
\def\unity{{\hbox{1\kern-.8mm l}}}

\newcommand{\opl}{\omega_{\rm pl}}
\newcommand{\LV}{{\rm erg }^{-1}\,{\rm cm}^{-3}\,{\rm s}^{-1} }
\newcommand{\den}{{\rm g}\,{\rm cm}^{-3} }
\newcommand{\Lg}{{\rm erg \,g}^{-1}\,{\rm s}^{-1} }

\newcommand{\pit}{\pi^{{T}}}

\newcommand{\ops}{{\it o}{\rm Ps}}

\renewcommand{\to}{\rightarrow}

\renewcommand{\to}{\rightarrow}

\newcommand\K{\mbox{K}}
\newcommand\eV{\mbox{eV}}
\newcommand\keV{\mbox{keV}}
\newcommand\MeV{\mbox{MeV}}
\newcommand\GeV{\mbox{GeV}}
\newcommand\TeV{\mbox{TeV}}

\def\nn{\\ \nonumber}
\def\de{\partial}

\begin{document}

\title{Extra dimensions, orthopositronium decay, and stellar cooling}
%

\author{A. Friedland}
\email{friedland@lanl.gov}
\author{M. Giannotti}
\email{maurizio@lanl.gov}

\affiliation{{\it Elementary Particles and Field Theory Group, MS B285, Los Alamos National Laboratory, Los
Alamos, NM 87545}}

\date{September 13, 2007}

\preprint{LA-UR-07-5609}


\begin{abstract}
  In a class of extra dimensional models with a warped metric and a
  single brane the photon can be localized on the brane by gravity
  only. An intriguing feature of these models is the possibility of
  the photon escaping into the extra dimensions. The search for this
  effect has motivated the present round of precision orthopositronium
  decay experiments. We point out that in this framework a photon in
  plasma should be metastable. We consider the astrophysical
  consequences of this observation, in particular, what it implies for
  the plasmon decay rate in globular cluster stars and for the
  core-collapse supernova cooling rate. The resulting bounds on the
  model parameter exceed the possible reach of orthopositronium
  experiments by many orders of magnitude.

\end{abstract}


 \maketitle

\section{Introduction}

Theories with extra dimensions
\cite{RS01,RS02,Visser1985,Antoniadis1990}\footnote{For an expanded
  set of references, see, {\it e.g.}, \cite{Visser1985,RubRev}.} have
been very popular in the last decade
\cite{ADD1,RSI,RSII,Gogberashvili}. In a large class of such theories
the extra-dimensional space is ``warped'', i.e. the metric scales
exponentially along one of the additional dimensions (see
Eq.~(\ref{eq:metric}) later). The scaling arises naturally as a
solution of Einstein's equations in the extra-dimensional ``bulk''
filled with a negative cosmological constant. This solution has the
same origin as the inflationary solution,
$a(t)\sim\exp(\sqrt{\Lambda}t/M_{pl})$, but is ``aligned'' to scale
along one of the spatial directions, rather than the time direction.
This ``inflation along a spatial direction'' can be arranged by
introducing one or more domain walls (branes), tuning their
tension(s), and replacing the positive cosmological constant of
inflation with a negative value. In a model with two branes, this
setup holds promise for solving the hierarchy problem, as the
exponents can reduce ratios of vastly different scales to relatively
modest numbers \cite{RSI}.
%
The same reasoning may explain the smallness of the Yukawa couplings
\cite{Yukawa}.  In a model with a single brane, this setup offers an
alternative to compactification \cite{RSII} and, through the AdS/CFT
(holographic \cite{holography}) connection \cite{Maldacena}, could in fact describe a
four-dimensional world with a new conformal sector.

Of course, in any such theory, one faces the problem of explaining why
we see only four space-time dimensions. One possible line of argument
is that the Standard Model fields could be \emph{dynamically} confined
to the four-dimensional Minkowsky defect (brane), as discussed already
two decades ago (RS0) \cite{RS01}.  In models with warping, it is
possible that the fields are localized to the brane \emph{by the
  metric itself}, {\it i.e.}, by gravity. The localization of the
graviton by this mechanism in the model with a single warped extra
dimension was discussed already in a seminal paper \cite{RSII} (RSII).
A scalar field can be similarly localized \cite{BajcGabadadze}. Gauge
fields are not localized in the minimal setup of RSII, but can be
localized if the model is extended with additional compact extra
dimensions \cite{RubCharge}.

The states localized in this way act most of the time as ``normal''
four-dimensional massless particles.  
Under some circumstances, however, they can tunnel into the extra
dimensions, ``disappearing'' from our world. The tunneling can happen
if the state is given a nonzero mass, or if it is produced as a virtual
state with time-like momentum. Not only is this suggestion intriguing,
but, more importantly, potentially experimentally testable.

We will focus on the possibility of photon tunneling. This effect
would manifest itself as unexplained missing energy events at an
$e^+e^-$ collider. The measurements of the $Z$ boson resonance
provide considerable bounds on the allowed curvature of the extra
dimension (see later). Another interesting experimental direction 
that is being actively pursued is the search for an invisible mode in
the orthopositronium decay \cite{Crivelli07}. The orthopositronium
serves as an $e^+e^-$ collider with a hermetic detector. Compared to
the $Z$ resonance measurement, one obviously loses on the
center-of-mass energy, but gains considerably on the sensitivity to the
branching ratio into the invisible mode. The recently published results~\cite{Crivelli07}
find the bound ${\rm Br}(\ops \to {\rm extra~ dim})\leq 4.2\times
10^{-7}$, with ${\rm Br}(\ops \to {\rm extra~ dim})\leq 10^{-8}-
10^{-9}$ expected in the future \cite{GKR}.

In this paper, we point out that in the same framework 
photons in plasma (plasmons) should also be subject to the invisible
decay. Indeed, plasma modifies the photon dispersion relation, in a
sense providing it with a mass, thereby opening up the decay channel.
In what follows, we consider the effects of the additional cooling on
the cores of low-mass red giants, horizontal branch stars, and
core-collapse supernovae. The bounds we find on the model parameter
exceed the possible reach of the orthopositronium by many orders of
magnitude.

\section{Tunneling into extra dimensions: overview}

As already mentioned, the existence of the photon mode localized on
the positive tension brane in the scenarios with warped extra
dimension(s) is well established. Following \cite{RubCharge}, let us
consider a space with the metric
\begin{equation}
\label{eq:metric}
        ds^2=a(z)^2 \left(  \eta_{\mu\nu}dx^\mu dx^\nu-\delta_{ij}d\theta_i d\theta_j\right)-dz^2\,.
\end{equation}
Here $z$ labels the {\it infinite} warped extra dimension,
$a(z)=\exp{(-k|z|)}$. At $z=0$ we have a domain wall (brane) with
positive tension. The variables $\theta_i\in[0,2\pi R_i]$ label $n\geq 1$
additional compact dimensions, with radii $R_i$. 
The fields are assumed to be independent of $\theta_i$.  

The action for the electromagnetic field $A_{C}(x,z)$ in this space is
\begin{eqnarray}
\label{action}
    S &=& \int\sqrt{|g|}\,d^4x\,dz  \prod_{i=1}^{n}
    \frac{d\theta_i}{2\pi\,R_i}  \, \mathcal{L} \,, 
 \\
 \label{eq:L_vac}
    \mathcal{L} &=& -\frac{\Lambda}{4} F_{CD} F^{CD}  \,.
\end{eqnarray}
In Eq. (\ref{action}), 
$F_{CD}\equiv\partial_{C}A_{D}-\partial_{D}A_{C}$
and $\Lambda$ is a
constant with mass dimension $1$, which will be determined later from
the requirement that the standard four-dimensional coupling is reproduced.
The Latin indices are assumed to run over all 
coordinates, including the extra dimensions.

The equations of motion in vacuum is $\de_C
(\sqrt{g}F^{CD})=0$. In the $A_5=0$ gauge this reads \cite{RubRev}
\begin{eqnarray}
  \label{eq:eqofmotion_mu}
     \eta^{\lambda\nu} \de_\lambda F_{\nu\mu} &=& 
     \de_z(a(z)^{n+2}\de_z A_\mu)/a(z)^{n},\\
  \label{eq:eqofmotion_z}
     \de_z(\eta^{\mu\nu}\de_\mu A_\nu) &=& 0.
\end{eqnarray}
The Greek indices run through 0, 1, 2, 3 (our space-time) and
$\eta^{\mu\nu}$ is the usual Minkowsky metric. As pointed out in
\cite{RubRev}, this system of equations has an obvious solution that
is independent of $z$, $A_\mu(x,z)\rightarrow A_\mu(x)$.
Eq.~(\ref{eq:eqofmotion_z}) is trivially satisfied in this case, while
Eq.~(\ref{eq:eqofmotion_mu}) becomes the usual Maxwell's equation for
a massless photon. This solution describes the \emph{zero mode}
localized on the brane. The reason this is so is because the
eigenfunctions are normalized with the integration measure $\int dz
a^n$ (hence the need to introduce the compact dimensions).

In general, the eigenfunctions are plane waves, $e^{-ipx}$, as a function of
$x=0,1,2,3$, owing to the fact that the Poincare invariance along the
brane is preserved.
The $z$ dependence of the eigenfunctions is given by the eigenmodes of
the operator on the right hand side of Eq.~(\ref{eq:eqofmotion_mu}).
Denoting the eigenvalue by $m^2$, we find that $m^2 = p^2$ and
\begin{equation}
\label{eq:antifriction}
        -\de_z^2 A_\mu(z) + (2+n)\,k\,{\rm sign} (z)\,\de_z
A_\mu(z)=e^{2 k |z|} m^2 A_\mu(z).
\end{equation}
From the four-dimensional point of view, the higher modes behave as
massive photons (Eq.~(\ref{eq:eqofmotion_mu}) for a given value of
$m^2\ne0$ takes the form of the Proca equation). As we will see
shortly, their eigenfunctions are strongly suppressed on the brane.

Further physical insight can be gained by recasting this equation in
the Schr\"odinger form. After changing the variable $z\to
s=sign(z)[\exp{(k|z|)}-1]$ and the redefinition of the fields as
$A_\mu(z) \to \phi_\mu(z)=A_\mu(z)\,\exp{[-k|z|(n+1)/2]}$ we get
\begin{equation}
\label{eq:Schrodinger_vacuum}
        \left[ -\frac12 \frac{\partial^2}{\partial s^2} + 
        \frac{(n+1)(n+3)}{8(|s|+1)^2}  
         - \frac{n+1}{2} \delta(s)   \right]\, \phi_\mu
        =\frac{m^2}{2\,k^2}\phi_{\mu}
\end{equation}
This transformation is similar to what was done in \cite{RSII} for the
graviton. Not only has the first derivative term disappeared, but also
the measure with which $\phi$ is normalized is trivial, $\int
ds$. Eq.~(\ref{eq:Schrodinger_vacuum}) thus describes a
non-relativistic Schr\"odinger problem and the usual physical
intuition fully applies here. We have a particle of unit mass in a
``volcano'' potential\footnote{Notice that the coefficients in
  Eq.~(\ref{eq:Schrodinger_vacuum}) are different from those for the
  graviton, which has a localized solution even for $n=0$.}, with a
confining $\delta$-function at the origin and a positive barrier
outside that slopes off to zero as $|s|\rightarrow\infty$.  This
potential can support a single bound state \emph{of zero energy} with
the wave function $\phi_0(s)=\sqrt{n/2}(1+|s|)^{-(n+1)/2}$, which corresponds
to the flat solution of the original equation.

It is clear in this picture that the spectrum of states residing away
from the origin starts from zero energy and is continuous. This means
that the localized state is only \emph{marginally bound}: an
infinitely small perturbation to this setup that lifts the zero mode,
$0\rightarrow E'\equiv m^2/2k^2$ (for example by decreasing in
absolute value the coefficient of the $\delta$-function) of the
localized state makes it metastable. The particle can then tunnel
through the potential barrier and escape from the brane \cite{RubRev}.
The eigenvalue in this case becomes complex and the eigenfunction at
$|z|\rightarrow\infty$ has an asymptotic form of outgoing plane waves.

The decay rate due to tunneling for this class of problems can be
estimated as follows. The turning points of the tunneling on either
side of the brane are given by the condition
$(n+1)(n+3)/8(|s_0|+1)^2=E'$. For $s\gtrsim|s_0|$ the solution
asymptotes to the plane wave, $a(E') e^{-i \sqrt{2E'}|s|}$, while for
$s\lesssim|s_0|$ it can be approximated by the unperturbed function,
$\sqrt{n/2}(1+|s|)^{-(n+1)/2}$. The amplitude of the plane wave $a(E')$
(the barrier penetration factor) can
then be estimated as roughly the unperturbed solution at the turning
point, $|a(E')|^2\sim (E')^{(n+1)/2}$. The flux away from the brane
computed at large $|s|$ equals $2|a(E')|^2 \sqrt{2E'} \sim
(E')^{(n+2)/2}$. The \emph{ratio} of the decay rate $\Gamma'$ to the
energy of the metastable state $E'$ is $\sim (E')^{n/2}$, true in any
system of units. In the normal units in which the energy is $m$, we
thus obtain for the decay rate in the rest frame \cite{Rub1}
\begin{equation}
  \label{eq:gamma}
  \Gamma_0^{vac} = c_n m(m/k)^n.
\end{equation}
The numerical coefficient $c_n$ can be found by considering the
properties of the exact solution given by the Hankel functions
\cite{Rub1}. We find
\begin{equation}
  \label{eq:c_n}
  c_n=(\pi n)/(2^{n+1}\Gamma[n/2+1]^2),
\end{equation}
where $\Gamma$ denotes the gamma function. Numerically,
$c_n=(1,\pi/4,1/3,\pi/32,1/45,...)$ for $n=(1,2,3,4,5,...)$.

We can also now easily see that the continuum modes residing in the
bulk are suppressed on the brane. Indeed, they have to tunnel to the
brane \emph{from the outside}. This suppresses the wave functions by
the barrier penetration factor $\sim(m/k)^{(n+1)/2}$, making the model
phenomenologically viable for energies $\ll k$.

Another way to describe the escape into the extra dimensions is by
inspecting the propagator between two points on the brane \cite{Rub1}.
As shown in \cite{Rub1}, the Fourier transform of this propagator for a
massless localized scalar in the Randall-Sundrum background ($n=0$) is
\begin{eqnarray}
  \label{eq:propagator}
  \left[\frac{p}{k}\frac{H_{1}^{(1)}(p/k)}{H_{2}^{(1)}(p/k)}  \right]^{-1}
  \approx
  2 k^2\left[p^2 + i \frac{\pi}{(\Gamma(2))^2}p^2\left(\frac{p}{2k}\right)^{2} \right]^{-1},
\end{eqnarray}
where $H^{(1)}$ denotes the Hankel function of the first kind and
$p\equiv\sqrt{p^2}$, where $p^2$ is the square of the four dimensional
momentum.  This approach makes it very clear that time-like virtual
particles are also subject to tunneling.  For $p^2>0$, up to the
overall normalization factor, this propagator has a standard
Breit-Wigner form $[p^2+ip\Gamma]^{-1}$ with the imaginary part giving
the decay rate, $\Gamma(\gamma^\ast\rightarrow \mbox{extra
  dim})=(\pi/4) \sqrt{p^2}(p^2/k^2)$.

The time-like virtual photon is formed, {\it e.g.}, in $e^+e^-$
annihilation at colliders. The bound from the measurements of the $Z$
width at LEP \cite{GKR} is $k\gtrsim m_Z (c_n m_Z/\Delta
\Gamma_Z^{inv})^{1/n}$, where $\Gamma_Z^{inv}<2.0$ MeV is the limit on
the additional invisible decay width of $Z$. For $n=2$ this yields
\cite{GKR} $k\gtrsim 17$ TeV. Clearly, getting a tighter bound on
$\Gamma^{inv}$ would improve the bound, which is the idea behind
looking for this process in orthopositronium decay. The invisible
width is $\Gamma (\ops \to {\rm extra~ dim}) \sim c_n m_{\ops}
(m_{\ops}/k)^n \alpha^4$ (one power of $\alpha$ comes from the photon
vertex, and three more from the wavefunction of $\ops$ at the origin),
compared to the standard three-photon width, $\Gamma(\ops\to
3\gamma)\sim m_{\ops}\alpha^6$ \cite{landavshitz}.  One gets a bound
$k\gtrsim m_{\ops} (c_n m_{\ops}/\Delta \Gamma_{\ops}^{inv})^{1/n}
\alpha^{4/n}=m_{\ops} (c_n/BR_{\ops}^{inv})^{1/n} \alpha^{-2/n}$.
Compared to the LEP bound, one trades a factor of $m_Z/m_{\ops}\sim
10^5$ for a factor of $(\Gamma_Z^{inv}/(m_Z\alpha^2
BR_{\ops}^{inv}))^{1/n}$.  Properly keeping track of all coefficients,
one finds \cite{GKR}, for $n=2$, $k\gsim 0.5\TeV$ with the present
accuracy $BR(\ops \to {\rm extra~ dim})\leq 4.2\times 10^{-7}$. If the
bounds on $BR(\ops \to {\rm extra~ dim})$ are improved to the
$10^{-10}$ level, the orthopositronium bound for $n=2$ would surpass
that of LEP. Note that for larger $n$ LEP has a bigger advantage, in
particular $k_{LEP}(n\to\infty)>m_Z$, while
$k_{\ops}(n\to\infty)>m_{\ops}$.

\section{Plasmon decay to extra dimension}

Let us now consider the effect of plasma on the zero mode. Strictly
speaking, one needs to specify how the electrons
in plasma are localized to our brane, for example with a domain wall
in a new scalar field. While the fine details will be model
dependent, the essential features can be obtained by assuming the
localization ``by hand'', with a delta function, in the spirit of
\cite{DubovskyRubakov2002}. 

The effective photon Lagrangian, Eq.~(\ref{eq:L_vac}), gains an
additional term, $-(1/2)A_C \Pi^{CD} A_D \delta(z)$, where
$\Pi^{CD}=\langle j_C j_D \rangle$ is the photon self-energy in
plasma, or truncated forward scattering matrix element
\cite{RaffBook}. The presence of this term changes the equation of
motion to $\Lambda\de_C (\sqrt{g}F^{CD})=-\Pi^{CD} A_C\delta(z)$.

Let us describe the main properties of $\Pi$.  First of all, since
$\Pi_{C5}=\Pi_{5C}=0$, it has a block diagonal form.  In addition we
can see from the above equations that $\Pi_{55}$ plays no role in our
gauge ($A_5=0$).  Let us therefore concentrate on the $4-$dimensional
part $\Pi_{\mu\nu}$.  In the hypothesis of isotropic plasma the tensor
$\Pi_{\mu\nu}$ is diagonalizable.  Because of the gauge invariance,
one of the eigenvectors is directed along the photon 4-momentum $q$,
and has zero eigenvalue. The others define the directions
$\epsilon^{(i)}$ of the different physical polarizations, and have in
general non-vanishing eigenvalues $\pi^{(i)}$.  If we assume parity
invariance the two {\it transverse} modes have the same eigenvalue
$\pit$, whereas the eigenvalue of the {\it longitudinal} mode,
$\pi^L$, is in general different.

In the basis spanned by $\epsilon^{(i)}$, the equation of motion is
diagonal. Eq.~(\ref{eq:Schrodinger_vacuum}) in the presence of plasma
generalizes to
\begin{equation}
\label{eq:Schrodinger_plasma}
        \left[ -\frac12 \frac{\partial^2}{\partial s^2} + 
        \frac{(n+1)(n+3)}{8(|s|+1)^2}  
         - \left(\frac{n+1}{2} -\frac{\pi^{(i)}}{2 k\Lambda} \right) \delta(s)   \right]\, \phi^{(i)}
        =\frac{m^2}{2\,k^2}\phi^{(i)},
\end{equation}
where $\phi^{(i)}$ denotes the components of $\phi$ along
$\epsilon^{(i)}$.  Treating the plasma term as a perturbation, in the
lowest order of perturbation theory we can write that the zero mode is
lifted by the energy $\delta E'=\langle \phi_0 | \pi^{(i)}
\delta(s)/(2 k\Lambda) |\phi_0\rangle=\pi^{(i)} (n/2)/(2 k\Lambda)$.
To reproduce the four-dimensional phenomenology, we write $\Lambda=k
n/2$, $m^2=\pi^{(i)}$.  The rest of the argument proceeds analogously
to the vacuum case considered earlier. The bound state becomes
metastable and the corresponding decay rate into extra dimensions in
the rest frame is given by
\begin{equation}
  \label{eq:gamma_plasma}
  \Gamma_0^{pl\;(i)} = c_n \sqrt{\pi^{(i)}}(\sqrt{\pi^{(i)}}/k)^n.
\end{equation}

The quantities $\pi^{(i)}$ are related to the plasma frequency,
$\sqrt{\pi^{(i)}}=\zeta^{(i)}\opl$. Here $\zeta^{(i)}$  is in
general a function of the photon energy and momentum. Fortunately, for 
transverse photons it can be shown \cite{Braaten, RaffBook} to be always
close to one,  
$1\leq \zeta^T \leq \sqrt{3/2}$. Moreover, the contribution of 
longitudinal photons to stellar cooling rates in all cases of
interest to us can be neglected.

Lastly, for the purpose of computing the cooling rate we need the
decay rate of a moving plasmon, $\Gamma_\omega^{pl\;(i)}$. The latter
is related to the one given in Eq.~(\ref{eq:gamma_plasma}) by the
Lorenz factor,
\begin{equation}
  \label{eq:ED}
  \Gamma_\omega^{pl\;(i)} =
\Gamma_0^{pl\;(i)}\sqrt{\pi^{(i)}}/\omega = c_n (\zeta^{(i)})^{n+2}\opl
(\opl/\omega)(\opl/k)^n,
\end{equation}
where $\omega$ is the energy of the plasmon.

\section{Implications}

\subsection{Astrophysical Bounds}

The energy loss rate per unit volume in a star is computed as (decay
rate) $\times$ (energy loss) $\times$ (photon number density), {\it
  i.e.}, as the integral of $\omega\Gamma_\omega^{pl\;(i)}$ over the
phase space ({\it e.g.}, \cite{RaffBook}),
\begin{equation}
\label{eq:Q}
 Q_T= \frac{\Gamma \, \omega}{\pi^2}(\zeta\, \opl)^3 \, g(\zeta\,\opl/T) \, ,
\end{equation}
where
$g(x)=\int_{1}^{\infty}(\xi\sqrt{\xi^2-1})/(\exp{(\xi\,x)}-1)d\xi$,
and the subscript $T$ reminds that this is the contribution from
transverse photons only.

Let us consider, first, the stars on the Red Giant (RG) branch.  For
RG stars (at the helium flash) the internal temperature is about
$T\simeq 10^8\K$ and the density $\rho\simeq 10^6 \den$.  In these
conditions, the main standard cooling mechanism is the plasmon decay
into neutrinos (see, e.g., \cite{Clayton, RaffBook}).  The rate for
this decay is
\begin{equation}
   \label{eq:SM}
   \Gamma_{{\rm SM}}^{(i)}= 
   \frac{1}{48 \, \pi^2\,\alpha}\frac{Z^{(i)}\, C_V^2\,G_F^2 \zeta^{(i)\, 6}\,\opl^{6} }{\omega}, 
\end{equation}
where $Z^{(i)}$ is a renormalization constant whose value is $\simeq
1$ for transverse photons and between $0$ and $1$ for longitudinal
photons \cite{Braaten}, $C_V=0.96$ is the vector-current coupling
constant and $G_F=1.166\times 10^{-5}\,\GeV^{-2}$ is the Fermi
constant. The contribution of longitudinal photons to this cooling is
always less than $10\%$ \cite{Haft} and we will neglect it in what
follows.  We will also neglect the longitudinal contribution to the
non-standard cooling, since this is certainly a conservative
assumption.

Stellar models with the cooling rate in Eq.~(\ref{eq:SM}) are in good
agreement with observations of globular cluster populations \cite{mu}.
To maintain this agreement, we need to constrain any additional energy
loss to not exceed about twice the standard neutrino luminosity
\cite{Haft}. From (\ref{eq:Q}), (\ref{eq:SM}) and (\ref{eq:ED}), we
find
\begin{eqnarray}
&&  \frac{ Q_{{\rm ED}} }{ Q_{{\rm SM}} }=\frac{ \Gamma_{{\rm ED}} }{ \Gamma_{{\rm SM}} }
=\frac{c_n\,(\zeta^T)^{n-4}}{u}   \left(  \frac{M_W}{\opl}  \right)^4\,  
     \left( \frac{ \opl }{k} \right)^{n} \nn
&&  \simeq 2.8\times 10^{30-8n}c_n  \left( \frac{ \zeta\,\opl }{10 \keV} \right)^{n-4}
\left(\frac{1\, \TeV}{k}\right)^{n}
\end{eqnarray}
where $u=(C_V^2g^4)/(1536\,\pi^2\alpha)\simeq 1.5\times 10^{-3}$,
$g\simeq0.65$ is the weak coupling constant, and we set $Z^{T}=1$.  If
we impose that this does not exceeds about $2$ we find
\begin{equation}
\label{Eloss}
k\geq \zeta^T c_n^{1/n}\,\opl\,B^{1/n}
\end{equation}
where for RG stars
\begin{equation}
\label{RG}
B= \frac{1}{2u\,(\zeta^T)^{4}} \left( \frac{ M_W }{\opl} \right)^{4}\, .
\end{equation}
In the nonrelativistic limit, the plasma frequency is given by
$\opl=28.7\eV (Y_e \rho)^{1/2}(1+(1.0\times 10^{-6}Y_e
\rho)^{2/3})^{-1/4}$ \cite{RaffBook}, where $\rho$ is in units of
$\den$ and $Y_e$ is the electron fraction.  To be conservative, we
take for $\zeta^T$ its largest value and for $\opl$ its value in the
center of the star just before helium flash, $\opl\simeq
17.8\keV$\cite{Haft}, corresponding to $\rho \simeq 10^6\den$.  This
choice leads to $B=6.2\times 10^{28}$ and to the bounds
\begin{eqnarray}
\label{boundsRG}
&&   k \gsim 183\,M_W\,\left( \frac{M_W}{\opl}  \right)^3 \simeq  1.4 \times 10^{21} \, \TeV \,, \quad (n=1) \,, \nonumber  \\
&&   k \gsim 13\,M_W\,\left( \frac{M_W}{\opl}  \right) \simeq  5 \times 10^6 \, \TeV \,, \quad (n=2) \,, \nonumber  \\
&&   k \gsim 4.5 \,M_W\,\left( \frac{M_W}{\opl}  \right)^{1/3} \simeq         60 \, \TeV \,, \,\quad (n=3) \,,
\end{eqnarray}
which are many orders of magnitude stronger than the direct laboratory
bounds\footnote{The bound for $n=1$ should be interpreted to mean that
  the model is excluded for all energies for which it is a valid
  effective description, possibly up to the Planck scale.}.


Next, let us consider the effects of the extra cooling on the
supernova SN1987A.  It is known that after the explosion this
anomalous energy loss cannot significantly exceed $Q_{{\rm Max}} =
3\times 10^{33} \LV$ ({\it e.g.}, \cite{RaffBook}), which corresponds
to the energy released in neutrinos.  The plasma frequency in a SN
core is approximately given by $\opl^2\simeq 4\alpha\mu^2/3\pi \sim
(10\MeV)^2$, where $\mu\sim 200 \MeV$ \cite{KeilJankaRaffelt1994} is
the electron chemical potential.  Therefore $\opl < T\simeq 30\MeV$,
in which case one can approximate $g(x)\simeq 2/x^3$, giving
$Q_T\simeq 2\Gamma\omega\,T^3/\pi^2$.  We find again a bound as in
Eq.~(\ref{Eloss}), but with
\begin{eqnarray}
\label{AB}
 B=\frac{2 (\zeta^T \opl)^2\,T^3}{\pi^2 Q_{Max}} \simeq  
   5.8\times 10^{19}\left(\frac{T}{30\MeV}\right)^3
\left(\frac{\opl}{10\MeV}\right)^2,
\end{eqnarray}
which implies 
\begin{eqnarray}
\label{boundsSN}
&&   k \gsim 6 \times 10^{14} \, \TeV \,, \quad (n=1) \,, \nn
&&   k \gsim 7 \times 10^4 \, \TeV \,, \quad (n=2) \,, \nn
&&   k \gsim 27 \, \TeV \,, ~~~\,\qquad (n=3) \,,
\end{eqnarray}
where, to be conservative, we have used $\zeta^T=1$.

Finally, a similar argument applies to stars on the horizontal branch
(HB).  Typically HB stars have an average temperature in the Helium
core of $T\simeq 0.8\times 10^8\K$ and density $\rho\simeq 0.5 \times
10^4\den$ \cite{Dearborn:1989he,RaffBook}. This implies, $\opl\simeq
1.5 \keV<T$.  In this case, an anomalous energy loss cannot be larger
than $Q_{ Max}/\rho\simeq 10$ $\Lg$ in order to have good agreement
between the predicted and observed number ratio of HB and RG stars
\cite{RaffBook}. In this case we find ($\zeta^T=1$)
$B=1.7\times10^{30}T_8^3 (\opl/1.5\keV)^2$, where $T_8=T/10^8\K$.  The
corresponding bounds are
\begin{eqnarray}
\label{boundsHB}
&&   k \gsim 1.1 \times 10^{21} \, \TeV \,, \quad (n=1) \,, \nn
&&   k \gsim 1.1 \times 10^6 \, \TeV \,, \quad (n=2) \,, \nn
&&   k \gsim 9 \, \TeV \,, ~~~\,\qquad (n=3) \,.
\end{eqnarray}

\subsection{Implications for Orthopositronium Decay}

The bounds we just found can be directly translated into the value of
the branching ratio (BR) necessary to have an analogous bound from the
orthopositronium experiment:
\begin{eqnarray}
\label{ops2}
     &&  {\rm BR}= \frac{\Gamma(\ops \to {\rm extra~ dim})}{\Gamma(\ops\to 3\gamma)}
        \simeq 1.5\times 10^5 c_n \left(\frac{m_{\ops}}{k}\right)^n  \nonumber   \nn
     && <\frac{1.5\times 10^5 }{B}  \left(\frac{m_{\ops}}{\opl}\right)^n  \,,
\end{eqnarray}
where, in the last step, we set $\zeta^T=1$ for simplicity.
Approximately this means $ {\rm BR}\lesssim 2\times 10^{-24+1.75n}$, from RG,
$ {\rm BR}\lesssim 2\times 10^{-15-1.48 n}$, from SN87A, $ {\rm BR}\lesssim 2\times
10^{-25+2.8 n}$, from HB stars.  Thus the astrophysical bounds on the
allowed branching ratio of $\ops$ to extra dimensions for $n=2$ are
some 14 orders of magnitude stronger than the present sensitivity of
the $\ops$ experiments. Moreover, the bound from supernova cooling is
at least 8 orders of magnitude more stringent than the present
experiments \emph{for any value of} $n$.

\section{Conclusions}

We have seen that the models in which the photon is gravitationally
trapped on the brane face significant constraints from astrophysical
considerations. The exact constraint depends on the number $n$ of
extra compact dimensions.  For $n=2$ or 3 the AdS curvature $k$ is
constrained to be orders of magnitude above the electroweak scale. For
$n=1$ the bound extends all the way to the Planck scale. For $n\ge 4$,
the astrophysical bounds are weaker than those coming from the LEP
measurement of the $Z$.  For any $n$, the astrophysical bounds imply
the rate of orthopositronium decay into extra dimensions that is at
least eight orders of magnitude smaller than the present experimental
sensitivity.  It is this implication for the ongoing and planned
experiments that provides the main motivation for our work.

A detailed discussion of the implications for the models is beyond the
scope of this paper. Briefly, our bounds do not exclude the models,
but provide significant constraints on them. One way to keep the
scales in the model close to the electroweak scale is by having a
large ($n\gtrsim 4$) number of extra dimensions. Another possibility
is to arrange for an additional binding mechanism for the photon,
besides gravity. The binding energy in the latter case needs to
significantly exceed the plasma frequency in the proto-neutron star
inside a core-collapse supernova ($\sim 10$ MeV).


\begin{acknowledgments}
We would like to thank Vincenzo Cirigliano and Elizabeth Price for helpful
comments.
\end{acknowledgments}


\end{document}